\documentclass[aps,prb,twocolumn, showpacs]{revtex4}

\usepackage{epsfig}
\usepackage{graphicx}

\def\be{\begin{equation}}
\def\ee{\end{equation}}
\def\bea{\begin{eqnarray}}
\def\eea{\end{eqnarray}}

\def\Xint#1{\mathchoice
   {\XXint\displaystyle\textstyle{#1}}%
   {\XXint\textstyle\scriptstyle{#1}}%
   {\XXint\scriptstyle\scriptscriptstyle{#1}}%
   {\XXint\scriptscriptstyle\scriptscriptstyle{#1}}%
   \!\int}
\def\XXint#1#2#3{{\setbox0=\hbox{$#1{#2#3}{\int}$}
     \vcenter{\hbox{$#2#3$}}\kern-.5\wd0}}

\def\dashint{\Xint-}

\begin{document}
\newcount\timehh  \newcount\timemm
\timehh=\time \divide\timehh by 60
\timemm=\time
\count255=\timehh\multiply\count255 by -60 \advance\timemm by \count255

\title{Spin currents, spin populations, and dielectric function}
\author{Emmanuel I. Rashba\cite{Rashba*} }
\affiliation{Department of Physics, MIT, Cambridge, Massachusetts 02139, USA}
\date{April 19, 2004}

\begin{abstract}
In Maxwellian electrodynamics, specific properties of the responses to external fields are included in constitutive equations. For noncentrosymmetric semiconductors, spin conductivity  can be expressed in terms of the contribution of electric-dipole transitions between spin-split spectrum branches to the dielectric function. In a dissipationless regime, a spin current driven by an external electric field is tantamount to a background current in an equilibrium system with a reduced symmetry. The importance of transients and gradients for efficient spin-current injection is emphasized.
\end{abstract}

\maketitle

The growing interest in using the electron (hole) spin degree of freedom in semiconductor spintronics \cite{Wolf} and impressing progress in the experimental study of spin-charge coupling through spin orbit (SO) interaction\cite{Silsbee} stimulated active research in nonequilibrium spin populations and spin currents in semiconductor microstructures. The problems with achieving efficient electrical spin injection from ferromagnetic electrodes, including the effect of stray magnetic fields around them, stimulated developing the concept of all-semiconductor spintronics that does not include any magnetic elements and is based on specially engineered spin injectors using SO coupling; a number of related ideas has been put forward.\cite{SOinj} An adiabatic pump for spin currents \cite{Marcus} and a Stern-Gerlach type experiment with a SO beam splitter\cite{Rokh} have been reported; both techniques require an external magnetic field. Independently, an  optical technique for a controlled injection of spin populations and spin currents has been proposed,\cite{BhSi} and injection of a pure spin current (with no net charge current and no net spin injected) has been reported.\cite{BhSiExp} The concept of spin-polarized currents also turned out highly productive for metals\cite{SlBer} and was applied to the spin-transfer induced switching of nanomagnets.

Several proposals for electrical injecting spin-currents into nonmagnetic materials that require neither magnetic fields nor ferromagnetic materials have been put forward recently. Governale {\it et al.} \cite{Gov} and Mal'shukov {\it et al.}\cite{Malsh} calculated spin currents driven by ac electric fields, while Murakami {\it et al.} \cite{Mur} and Sinova {\it et al.}\cite{Sino} proposed dissipationless spin currents driven by a dc electric field. The subject is of active interest, and an appropriate understanding of the nature of spin currents and their potentialities for spintronics is a challenging task.

The notion of spin current as {\it the transport of electron spins} in a real space sounds alien to the standard Maxwellian electrodynamics; the ``magnetization current" $c~{\rm curl}\mbox{\boldmath$M$}$ has a different nature and is a part of the charge current. In what follows, I establish a relation between the spin conductivity and the dielectric function. To keep the calculations and the results as easy as possible, I restrict myself by free electrons. Such an approach allows to pinpoint the specific details of the band structure and specific electronic transitions that are responsible for spin currents, and discuss the essence of these currents in the framework of the standard band theory. It also allows to identify the electric-field driven dissipationless spin currents\cite{Mur,Sino} as the {\it equilibrium background spin currents}\cite{R03} that develop in the system when it is subjected to a proper pyroelectric deformation.

Because spin currents are not conserved in systems with SO interaction, their definition is somewhat arbitrary. I apply the generally accepted and physically appealing definition of the SC pseudotensor ${\cal J}_{ij}$
\begin{equation}
{\cal J}_{ij}={1\over 2}\sum_\lambda\int{{d^2k}\over(2\pi)^2}\langle\lambda\vert\sigma_iv_j(\mbox{\boldmath$k$})+v_j(\mbox{\boldmath$k$})\sigma_i\vert\lambda\rangle,
\label{eq1}
\end{equation}
establish its relation to the dielectric function, and discuss the basic conditions for the generation of {\it transport spin currents}. Here \mbox{\boldmath$k$} and \mbox{\boldmath$v$} are the electron momentum and the velocity operator, respectively, \mbox{\boldmath$\sigma$} is the vector of Pauli matrices, $i,j$ are Cartesian coordinates, with $i$ indicating the spin component and $j$ the transport direction, and $\lambda$ numerates the spectrum branches. For $T=0$, the integration should be performed inside the Fermi surface. It is an important property of the crystals lacking the inversion center that SO interaction splits every energy band into two branches. Below, calculations are performed for a SO-split band of two-dimensional (2D) electrons with a Rashba SO interaction.\cite{Rash} A similar approach to 3D Luttinger holes\cite{Lutt} in a diamond type semiconductor will be also discussed.

In what follows, (i) frequency dependencies of the dielectric function $\epsilon(\omega)$ and spin conductivity $\Sigma(\omega)$ will be found, (ii) a relation between them established, and (iii) it will be shown that $\Sigma(\omega)$ is directly related to the contribution to the real part of $\epsilon(\omega)$ coming from the region of the \mbox{\boldmath$k$}-space where the lower branch is populated and the upper one is empty. 

The standard Hamiltonian of a $2\times 2$ SO problem is
\begin{equation}
H_R=\hbar^2k^2/2m+\alpha(\mbox{\boldmath$\sigma$}\times\mbox{\boldmath$k$})\cdot\hat{\bf z},
\label{eq2}
\end{equation}
with $\mbox{\boldmath$k$}=(k_x, k_y)$ the 2D momentum, $\hat{\bf z}$ a unit vector perpendicular to the confinement plane, and $\alpha$ the SO coupling constant. The eigenvalues of the Hamiltonian $H_R$ are 
$\varepsilon_\lambda(k)\equiv\hbar\omega_\lambda(k)=\hbar^2k^2/2m+\lambda\alpha k$, where $\lambda=\pm 1$ correspond to the upper and lower branches of the spectrum, respectively; $\alpha>0$. The eigenspinors are
\begin{equation}
\psi_\lambda (\mbox{\boldmath$k$})= {1\over \sqrt{2}}\left(
\begin{array}{c}
1\\-i\lambda(k_x+ik_y)/k
\end{array}\right),
\label{eq3}
\end{equation}
and the velocity operator is
\be
\mbox{\boldmath$v$}=\hbar^{-1}\partial H_R/\partial\mbox{\boldmath$k$}=
\hbar\mbox{\boldmath$k$}/m+\alpha({\hat{\bf z}}\times\mbox{\boldmath$\sigma$})/\hbar.
\label{eq4}
\ee

The electric current driven by an homogeneous electric field $\mbox{\boldmath$E$}e^{-i\omega t}$ (with a vector-potential $\mbox{\boldmath$A$}(t)=e\mbox{\boldmath$E$}(t)/i\omega$), according to Kubo formula, is expressed through a retired commutator of the currents $\hat{j}_i=ev_i$ 
\bea
j_i(t)&=&j_i^{(0)}(t)+j_i^{(1)}(t),\;\;\mbox{\boldmath$j$}^{(0)}(t)=-(ne^2/mc)\mbox{\boldmath$A$}(t),\nonumber\\
j_i^{(1)}(t)&=&{i\over{\hbar c}}\int_{-\infty}^t
\langle[\hat{j}_i(\mbox{\boldmath$k$},t),\hat{j}_j(\mbox{\boldmath$k$},t')]_-\rangle A_j(t')dt',
\label{eq5}
\eea
$n$ being the 2D election concentration, and the angle bracket $\langle ... \rangle$ indicates integration over the Fermi distribution in the $(\lambda,\mbox{\boldmath$k$})$ space. Because of the isotropy of the problem, this integral is diagonal in the indeces $(i,j)$. All matrix elements of $\hat{\mbox{\boldmath$j$}}=e\mbox{\boldmath$v$}$  diagonal in $\lambda$ cancel, and electric conductivity equals\cite{Magarill}
\bea
\sigma^{(1)}(\omega)&=&{{ie^2}\over{2\hbar\omega}}\int_{k_+}^{k_-}{{d^2k}\over{(2\pi)^2}}\langle-\vert\mbox{\boldmath$v$}(\mbox{\boldmath$k$})\vert+\rangle
\langle+\vert\mbox{\boldmath$v$}(\mbox{\boldmath$k$})\vert-\rangle
\nonumber\\
&\times&\biggl[{1\over{\omega_{-+}(\mbox{\boldmath$k$})+\omega+i\delta}}-{1\over{\omega_{+-}(\mbox{\boldmath$k$})+\omega+i\delta}}\biggr],
\label{eq6}
\eea
with $\omega_{\lambda\lambda^\prime}(\mbox{\boldmath$k$})\equiv\omega_{\lambda}(\mbox{\boldmath$k$})-\omega_{\lambda^\prime}(\mbox{\boldmath$k$})$ and $k_\pm$ indicating the integration limits. When the Fermi energy is positive, $\mu>0$, $k_+$ and $k_-$ are the Fermi radii for the upper- and lower-branch electrons, respectively; below, all equations are presented for this case. Hence, this contribution to the conductivity comes from the interbranch transitions in the $k$-space area where the lower branch is populated and the upper one is empty. Using the relation 
\be
\langle+\vert\mbox{\boldmath$v$}(\mbox{\boldmath$k$})\vert-\rangle=-
\langle-\vert\mbox{\boldmath$v$}(\mbox{\boldmath$k$})\vert+\rangle=(i\alpha/\hbar k)(\hat{\bf z}\times\mbox{\boldmath$k$})
\label{eq7}
\ee
and the equation $\epsilon(\omega)=4\pi i\sigma(\omega)/\omega$, one comes to the following equations for the real and imaginary parts, $\epsilon'(\omega)$ and $\epsilon''(\omega)$, of the 2D dielectric function
\bea
\epsilon'(\omega)&=&\epsilon'_{SO}(\omega)-\omega_p^2/\omega^2,\nonumber\\
\epsilon'_{SO}(\omega)&=&
4~{{e^2}\over\hbar}{{\alpha^3}\over{\hbar^3\omega^2}}\int_{k_+}^{k_-}{{k^2\;dk}\over{(2\alpha k/\hbar)^2-\omega^2}},
\label{eq8}
\eea
\be
\epsilon''(\omega)=\pi e^2/4\hbar\omega,\;\;{\rm when} \;\;2\alpha k_+\leq\hbar\omega\leq2\alpha k_-.
\label{eq9}
\ee
The first term in $\epsilon'(\omega)$ came from $\sigma^{(1)}(\omega)$ and the second from the $\mbox{\boldmath$j$}^{(0)}$-current of Eq.~(\ref{eq5}), $\omega_p^2=4\pi e^2 n/m$, and $\epsilon''(\omega)\neq0$ only in the frequency range of the interbranch transitions. Eq.~(\ref{eq9}) can be also found by {\it the Golden Rule}. The total oscillator strength of the interbranch transition equals $f_{SO}=(m\alpha/\hbar^2)^2/2\pi$.

It is instructive to compare Eqs.~(\ref{eq8}) and (\ref{eq9}) with the Kramers-Kronig relation
\be
\epsilon'(\omega)={2\over\pi}\dashint_0^\infty{{\omega'\epsilon''(\omega')}\over{(\omega')^2-\omega^2}}d\omega';
\label{eq10}
\ee
a background dielectric function that is not related to the electron band in question is omitted in Eq.~(\ref{eq10}). If to define the Kramers-Kronig transform of Eq.~(\ref{eq9})
\be
\epsilon'_{KK}(\omega)={{e^2}\over{2\hbar}}\int^{2\alpha k_-/\hbar}_{2\alpha k_+/\hbar}{{d\omega'}\over{(\omega')^2-\omega^2}}
\label{eq11}
\ee
and employ the relation $k_--k_+=2m\alpha/\hbar^2$, then Eq.~(\ref{eq8}) can be rewritten as
\bea
\epsilon'(\omega)&=&\epsilon'_{KK}(\omega)-{{{\bar\omega}_p^2}/{\omega^2}},\nonumber\\
{\bar\omega}_p^2&=&\omega_p^2\left[1-{{(m\alpha/\hbar^2)^2}/{2\pi n}}\right],
\label{eq12}
\eea
the ratio $m\alpha/\hbar^2$ being the characteristic SO momentum. To conform Eqs.~(\ref{eq8}) and (\ref{eq10}) one needs to supplement $\epsilon''(\omega)$ of Eq.~(\ref{eq9}) with a singular contribution
\be
\epsilon''_{\rm sing}(\omega)=(\pi{{\bar\omega}_p^2}/2\omega)\delta(\omega).
\label{eq13}
\ee
This term reflects the oscillator strength that is hidden for free electrons and manifests itself in the cyclotron resonance, Drude absorption, etc. Eq.~(12) includes the renormalization of this oscillator strength by the SO coupling, ${\bar\omega}_p^2<\omega_p^2$. It is reduced because a part of it was borrowed for the interbranch absorption of Eq.~(\ref{eq9}).

To summarize, the function $\epsilon'_{SO}(\omega)$ describes the total contribution of the SO coupling into dielectric polarizability. It includes both the direct contribution from the interbranch transitions near the edge of the Fermi distribution and the reduction of the $\delta(\omega)$-part of $\epsilon''(\omega)$ because of the oscillator strength conservation. 

Spin current ${\cal J}_{zx}(t)$ driven by an electric field $\mbox{\boldmath$E$}(t)\parallel{\hat{\bf y}}$ can be calculated similarly to Eq.~(\ref{eq5}) by using ${\hat{\cal J}}_{zx}=(\hbar k_x/m)\sigma_z$. The spin conductivity equals
\bea
\Sigma_{zx}(\omega)&=&{{ie}\over{\hbar\omega}}\int_{k_+}^{k_-}{{d^2k}\over{(2\pi)^2}}\biggl[{{\langle+\vert v_y(\mbox{\boldmath$k$})\vert-\rangle\langle-\vert{\hat{\cal J}}_{zx}(\mbox{\boldmath$k$})\vert+\rangle}\over{\omega_{-+}(\mbox{\boldmath$k$})+\omega+i\delta}}\nonumber\\
&-&{{\langle+\vert {\hat{\cal J}}_{zx}(\mbox{\boldmath$k$})\vert-\rangle\langle-\vert v_y(\mbox{\boldmath$k$})\vert+\rangle}\over{\omega_{+-}(\mbox{\boldmath$k$})+\omega+i\delta}}\biggl].
\label{eq14}
\eea
Matrix elements $\langle\lambda\vert v_y(\mbox{\boldmath$k$})\vert\lambda'\rangle$, $\lambda\neq\lambda'$, are odd with respect to the transposition of $\lambda$ and $\lambda'$, while matrix elements $\langle+\vert\sigma_z\vert-\rangle=\langle-\vert\sigma_z\vert+\rangle=1$ entering into $\langle\lambda\vert {\hat{\cal J}}_{zx}(\mbox{\boldmath$k$})\vert\lambda'\rangle$ are even. As a result, in a factor similar to the bracket of Eq.~(\ref{eq6}) both fractions appear with the same sign. Finally, the real part of $\Sigma_{zx}(\omega)$ equals
\be
\Sigma'_{zx}(\omega)={{e\alpha}\over{2\pi\hbar m}}\int_{k_+}^{k_-}{{k^2\;dk}\over{(2\alpha k/\hbar)^2-\omega^2}}
\label{eq15}
\ee
in agreement with Schliemann and Loss.\cite{Loss} In the low-frequency limit, 
\be
\Sigma'_{zx}(\omega=0)=e/4\pi\hbar
\label{eq16}
\ee
 in agreement with Sinova {\it et al.}\cite{Sino} (after the difference by the factor $\hbar/2$ in the definition of ${\cal J}_{ij}$ is allowed for).

The comparison of Eq.~(\ref{eq8}) and (\ref{eq15}) shows that the integrals coincide and
\be
\Sigma'_{zx}(\omega)={\omega\over{8\pi e}}{{\hbar\omega}\over{m\alpha^2/\hbar^2}}\epsilon'_{SO}(\omega).
\label{eq17}
\ee
Hence, the spectrum-specific frequency dependence cancels from the ratio of $\Sigma'_{zx}(\omega)$ and $\epsilon'_{SO}(\omega)$. Therefore, spin conductivity $\Sigma'_{zx}(\omega)$, when properly normalized, acquires the meaning of the electron polarizability related to the transitions between the SO-split spectrum branches. Remarkably, the singular low-$\omega$ part of $\epsilon_{SO}'(\omega)$ corresponding to $\Sigma_{zx}'(\omega=0)$ comes from the SO correction to $\omega_p^2$, Eq.~(\ref{eq12}).

We conclude that spin currents and the SO part of the dipole moment, $\mbox{\boldmath$P$}_{SO}(t)=\epsilon_{SO}'(\omega)\mbox{\boldmath$E$}(t)$, represent two aspects of the same phenomenon.\cite{dipole} {\it Electrically driven spin currents are described macroscopically through the SO contribution to the dielectric function}.

The implications of this observation reveal themselves if one takes into account that in deriving Eqs.~(\ref{eq8}) and (\ref{eq15}) only the interbranch matrix elements of the perturbation $-(e/c)v_yA_y(t)$ were involved, while the effect of the electric field on the intrabranch motion dropped out. This is an artifact of applying Kubo formalism to free electrons, and doing so is an equivalent of the perturbation theory in the operator $V=-eEy=ieE\partial/\partial{k_y}$ in the $2\times2$ space of the spinors $\psi_\lambda(\mbox{\boldmath$k$})$. The linear in $V$ corrections to the spinors $\psi_\lambda(\mbox{\boldmath$k$})$ are
\be
\psi_\lambda^{(1)}(\mbox{\boldmath$k$})=-\lambda(eEk_x/4\alpha k^3)\psi_{-\lambda}(\mbox{\boldmath$k$}).
\label{eq18}
\ee
One can now find the mean value of ${\hat{\cal J}}_{zx}$ over the new vacuum spanned by the spinors $\psi_\lambda(\mbox{\boldmath$k$})+\psi_\lambda^{(1)}(\mbox{\boldmath$k$})$. The contributions from the upper and lower branches cancel, as before, in the region $k<k_+$. The contribution from the region $k_+<k<k_-$ equals ${\cal J}_{zx}=eE/4\pi\hbar$, in agreement with Eq.~(\ref{eq16}). This result corroborates that the Kubo approach, when applied to free electrons, allows for the effect of the field $\mbox{\boldmath$E$}(t)$ only through the rotation of spinors $\psi_\lambda(\mbox{\boldmath$k$})$. 

Such physics corresponds to a pyroelectric deformation along the $y$-axis rather to the effect of a transport electric field $\mbox{\boldmath$E$}\parallel{\hat{\bf y}}$. This pyroelectric field lowers the symmetry from the group $\mbox{\boldmath$C$}_{\infty v}$ of the Hamiltonian $H_R$ to the group $\mbox{\boldmath$C$}_s$ whose only nontrivial element is a reflection in the $yz$ plane. There are two new SO invariants, $\sigma_xk_y+\sigma_yk_x$ and $\sigma_zk_x$, in this group. The first can be disregarded, while the perturbation $H'_{SO}=\alpha_z\sigma_zk_x$ with $\alpha_z\approx-eE/2k_F^2$ results in spinors of Eq.~(\ref{eq18}) and the same spin current ${\cal J}_{zx}=eE/4\pi\hbar$ (when $k_--k_+\ll k_F$). Because a system with the Hamiltonian $H_R+H'_{SO}$ is in equilibrium, these spin currents are background (nontransport) currents.  

It follows from general arguments that {\it time-inversion symmetry forbids spin accumulation in a dissipationless dc regime}. Indeed, a linear relation between the magnetization $\mbox{\boldmath$M$}$ (or spin $\mbox{\boldmath$\sigma$}$) and $\mbox{\boldmath$E$}$ is equivalent to a magneto-electric effect\cite{Dzy} that is generally forbidden because $\mbox{\boldmath$E$}$ is real while $\mbox{\boldmath$M$}$ is imaginary with respect to $t$-inversion. For the magneto-electric effect to exist, this symmetry should be violated by a proper magnetic structure,\cite{Dzy} electron scattering,\cite{Levit} or a finite frequency.$\omega$\cite{Rash} Remarkably, spin polarization of free electrons that develops in the direction $(\mbox{\boldmath$E$}\times{\hat{\bf z}})\parallel{\hat{\bf x}}$ and diverges as $\omega^{-1}$ when $\omega\rightarrow0$, is cut-off by the momentum relaxation time $\tau_p$;\cite{Edel} this effect has been observed recently.\cite{AwGa}

In noncentrosymmetric crystals, the existence of equilibrium background spin currents\cite{R03,Pareek} is compatible with the $t\rightarrow-t$ symmetry. In fact, this symmetry requires that momenta and spins be reversed simultaneously, hence, it does not require that the currents of the particles with a given spin vanish.\cite{LifPit} Therefore,  {\it in a dissipationless regime Kubo formalism maps the real system driven by an electric field \mbox{\boldmath$E$} onto an auxiliary equilibrium system of a lower symmetry}. Both systems are described by identical equations because intrabranch dynamics has been eliminated. Spin currents flowing in the mimicking system are background currents. The optical analogy suggests that they set an upper bond for dc transport spin currents in the real system because the substitution $\omega\rightarrow-i\Gamma$ in Eq.~(\ref{eq15}), $\Gamma$ being a proper decay constant, results in a decrease of $\Sigma'_{zx}(\Gamma)$ with $\Gamma$.

The mapping of the real system onto an auxiliary equilibrium system is helpful because the problem of dissipationless currents cannot be posed rigorously in the absence of a strong magnetic field, and the mapping clarifies the assumptions underlying it.

Background spin currents do exist because the operator ${\hat{\cal J}}_{ij}$ is real with respect to $t$-inversion. For the same reason $\Sigma_{ij}'(\omega)$ is related to $\epsilon'(\omega)$, the dispersive part of $\epsilon(\omega)$, rather than to its dissipative part $\epsilon''(\omega)$. These currents are a reality. Nevertheless, they ``do not work" as spin sources because a pyroelectric element of a circuit cannot inject spins at equilibrium. To realize how the currents of this sort can be put to work, it is instructive to consider a simple classical analogy. The momentum flow $\Pi_{ii}=\Sigma_{\it l}v_i({\it l})p_i({\it l})$ is also real with respect to $t$-inversion, $\it l$ numerates particles inside a unit volume. For an equilibrium gas $\Pi_{ii}=P$, the pressure. In macroscopic terms, the equation for the momentum flow of an ideal fluid is $\partial(\rho v_i)/\partial t=-\partial\Pi_{ij}/\partial x_j$, where $\Pi_{ij}=P\delta_{ij}+\rho v_iv_j$ is the tensor of the momentum flow density, $\rho$ is a density.\cite{hydro} It is a gradient of $\Pi_{ij}$ that produces an acceleration, a flow, and transients. A similar approach is valid for spin currents. E.g., Mal'shukov {\it et al.}\cite{Malsh} have shown that modulating $\alpha=\alpha(t)$ results in injecting spin currents. In a diffusive regime they are controlled by the ratio $\omega/\Gamma$.\cite{Malsh} For abrupt changes in $\alpha$, ballistic pulses can be anticipated. In Stevens {\it et al.}\cite{BhSiExp} experiments, spin-current pulses spread from a small spot where they were generated by a laser pump.

Dynamics and propagation of spin populations and spin currents should be based on a theory including dissipation, and different approaches to this problem have already been advanced.\cite{Malsh,Loss,Burkov,Inoue,numeric} In particular, such a theory should provide generalizations of Eq.~(\ref{eq17}). When SO coupling is strong, $\alpha k_F\sim\mu$, the only characteristic time is $\tau_p$, the momentum relaxation time. Long relaxation times $\tau_S$ of spin populations,\cite{KikkAw} can be achieved when SO coupling is weak, $\alpha k_F\ll\mu$, and $\tau_p$ is short, $\tau_p\ll\hbar/\alpha k_F$. Then $\tau_S$, controlled by the Dyakonov-Perel process,\cite{DP} is long, $\tau_S^{-1}\approx\tau_p(2\alpha k_F/\hbar)^2$. In this regime, a quasiequilibrium in orbital degrees of freedom is established,\cite{DP,MR} and spin dynamics in an external electric field is strongly influenced by momentum scattering.\cite{MR} In the absence of spin populations, spin currents decay at the $\tau_p$ scale. Some of the emerging problems are similar to those of the physics of spin photocurrents.\cite{photo}

Holes in centrosymmetric crystals of the diamond type, described by a Luttinger Hamiltonian\cite{Lutt} 
\be
H_L=[(\gamma_1+5\gamma_2/2)k^2-2\gamma_2(\mbox{\boldmath$J$}\cdot\mbox{\boldmath$k$})^2]/2m,
\label{eq19}
\ee
are in some aspects similar to the electrons of the Hamiltonian $H_R$. Their spectrum consists of two (twice degenerate) branches known as heavy and light holes with the energies $(\gamma_1\mp2\gamma_2)k^2/2m$. The momentum \mbox{\boldmath$J$} is described by $4\times4$ matrices of the angular momentum $J=3/2$. The currents ${\cal J}_{ij}$ defined similarly to Eq.~(\ref{eq1}) with $\mbox{\boldmath$\sigma$}\rightarrow\mbox{\boldmath$J$}$ have the meaning of angular momentum currents and are related to interbranch transitions, in this case from the heavy hole to the light hole branch.\cite{indeed} Therefore, most of the above conclusions are applicable to this system. The main difference is related to the fact that for Luttinger holes the notion of ``spin" should be understood generally, as a total angular momentum rather than the physical spin, the constant $\gamma_2$ is only weakly influenced by the physical SO coupling,\cite{Lutt} and the heavy-light hole relaxation time is never long.

In conclusion, spin currents in noncentrosymmetric semiconductors are accompanied by electric dipoles related to electronic transitions between spin-split spectrum branches, and the polarization of eigenspinors by an electric field produces background spin currents. From this standpoint, the importance of transients and gradients for efficient electrical spin-current injection has been clarified.

Funding of this research granted through a DARPA contract is gratefully acknowledged.


\begin{thebibliography}{99}
\bibitem{indeed} The denominators $(\omega^2-4\gamma_2^2k^4/m^2)$ of Eq.~(26), Ref.~\onlinecite{Mur}b, and of Eq.~(\ref{eq8}) actually coincide because $2\gamma_2k^2/m$ equals the energy difference between the heavy and light holes.
\end{thebibliography}
\end{document}